\documentclass[conference]{IEEEtran}
\usepackage[pdftex]{graphicx}
\graphicspath{{_img/}}
\DeclareGraphicsExtensions{.jpeg,.png}

\usepackage[hyphens]{url}
\begin{document}
\title{Machine Learning with Memristors via Thermodynamic RAM}

\author{
  \IEEEauthorblockN{Timothy W. Molter}
  \IEEEauthorblockA{Knowm Inc., Santa Fe, NM, USA\\
  Email: tim@knowm.org}
  \and
  \IEEEauthorblockN{M. Alexander Nugent}
  \IEEEauthorblockA{Knowm Inc., Santa Fe, NM, USA\\
  Email: alex@knowm.org}
}

\maketitle

% abstract
\begin{abstract}
Thermodynamic RAM (kT-RAM) is a neuromemristive co-processor design based on the theory of AHaH Computing and implemented via CMOS and memristors. The co-processor is a 2-D array of differential memristor pairs (synapses) that can be selectively coupled together (neurons) via the digital bit addressing of the underlying CMOS RAM circuitry. The chip is designed to plug into existing digital computers and be interacted with via a simple instruction set. Anti-Hebbian and Hebbian (AHaH) computing forms the theoretical framework from which a nature-inspired type of computing architecture is built where, unlike von Neumann architectures, memory and processor are physically combined for synaptic operations. Through exploitation of AHaH attractor states, memristor-based circuits converge to attractor basins that represents machine learning solutions such as unsupervised feature learning, supervised classification and anomaly detection. Because kT-RAM eliminates the need to shuttle bits back and forth between memory and processor and can operate at very low voltage levels, it can significantly surpass CPU, GPU, and FPGA performance for synaptic integration and learning operations. Here, we present a memristor technology developed for use in kT-RAM, in particular bi-directional incremental adaptation of conductance via short low-voltage (\textless 1.0~V, \textless 1.0~$\mu$S) pulses.
\end{abstract}

\IEEEpeerreviewmaketitle

\section{Introduction}

Machine learning applications span a very diverse landscape. Some areas include motor control, combinatorial search and optimization, clustering, prediction, anomaly detection, classification, regression, natural language processing, planning and inference. A common thread is that a system learns the patterns and structure of the data in its environment by building a model, and then uses that model to make predictions of subsequent events and take actions. The models that emerge contain hundreds to trillions of continuously adaptive parameters. The human brain contain on the order of $ 10^{15} $ adaptive synapses. How the adaptive variables are exactly implemented in an algorithm varies, and established methods include support vector machines, decision trees, and artificial neural networks to name a few. Intuition tells us learning and modeling the environment is a valid approach in general, and the biological brain also appears to operate in this manner.

An unfortunate limitation when combining large-scale adaptive learning with computing is the memory-processing duality. In a traditional digital computer, calculations over adaptive variables must necessarily be performed in different physical locations than where the variables' memory resides, and this can often be a significant distance. The power required to adapt parameters grows impractically large as the number of parameters increases owing to the tremendous energy consumed shuttling digital bits back and forth between memory and processor. In a biological brain (and all of Nature), the processor and memory are the same physical substrate and computations and memory adaptations are performed in parallel. A truly low-power solution to machine learning occurs when the memory-processor distance goes to zero, and this can be achieved through intrinsically adaptive hardware such as memristors.

kT-RAM was designed with a few key constraints in mind, which are different than existing approaches: (1) must result in a hardware solution where memory and computation are combined, (2) must not only solve a single problem but provide a generic solution to most machine learning applications and (3) must be simple to fabricate with existing or emerging manufacturing technology. This initial motivation led us to create a theoretical framework for a neuromorphic processor satisfying the above constraints, which we call AHaH computing \cite{nugent2014ahah}. Independent works such as the Local Activity Principle \cite{mainzer2013cause} and Thermodynamics of Economics Modeling \cite{jorgensen2011evolutionary}, while more general theoretical frameworks compared to AHaH Computing, are based on roughly the same ideas of homogeneous matter spontaneously forming complex structure in order to maximize energy dissipation (4th Law of Thermodynamics).

Whereas previous publications present both AHaH Computing and Thermodynamic RAM in detail \cite{nugent2014ahah,nugent2014thermodynamic,nugent2014cortical}, here we will focus on the memristors we have designed and plan to use in the first kT-RAM prototype chips. 

\section{Results and Discussion}

We have recently developed and made commercially-available memristors with bi-directional incremental learning capability. This advancement opens the gateway to extremely efficient and powerful machine learning and artificial intelligence applications. The memristor devices come in three variants: W, Sn and Cr, which refers to the metal introduced in the active layer during fabrication resulting in different electrical characteristics. The memristor devices operate primarily through the mechanism of electric field induced generation and movement of metal ions through a multilayer chalcogenide material stack. The memristor devices are fabricated with a layer of metal that is easily oxidizable, located near one electrode. When a voltage is applied across the device with the more positive potential on the electrode near this metal layer, the metal is oxidized to form ions. Once formed, the ions move through the device towards the lower potential electrode. The ions move through a layer of amorphous chalcogenide material (the active layer) to reach the lower potential electrode where they are reduced to their metallic form and eventually form a conductive pathway between both electrodes that spans the active material layer, lowering the device resistance. Reversing the direction of the applied potential causes the conductive channel to dissolve and the device resistance to increase. The devices are bipolar, cycling between high and low resistance values by switching the polarity of the applied potential. The resistance is related at any time to the amount of metal located within the active layer. A secondary phase change mechanism can be used beforehand to initialize the memristor devices into the desired high and low conductance range.

\begin{figure}[!t]
\centering
\includegraphics[width=1.0\linewidth]{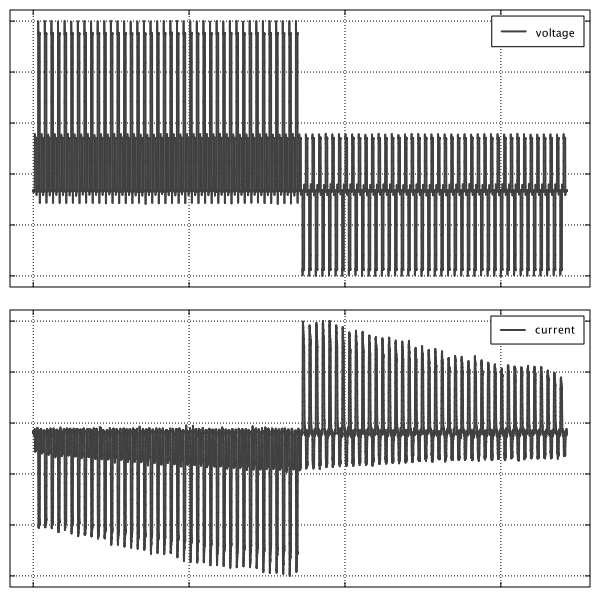}
\caption{A Tungsten Ag-chalcogenide memristor device exhibits bi-directional incremental learning behavior in response to 50~ns voltage pulses; useful for analog-based learning architectures such as Thermodynamic RAM. The voltage plot shows asymmetric forward and reverse activation pulses as well as lower-magnitude read pulses. The current plot shows the corresponding incremental change in conductance of the device occurring in both increasing and decreasing directions.}
\label{fig_memristor_pulsed}
\end{figure}

For bi-directional pulsed incremental learning applications, such as in kT-RAM, the primary mode of operation is the fine-grained metal ion transport within the device (Figure \ref{fig_memristor_pulsed}). Given one or more low-voltage pulses (\textless 1.0~V, \textless 1.0~$\mu$S), the memristor's conductance can be ``nudged'' into either a higher or lower value, representing a synaptic weight for example. In kT-RAM synapses are realized with two serially-connected memristors and used as a voltage divider where the output voltage is the synapse's weight. A read pulse across the memristor pair provides a ``weight'' and drives the synapse towards a forgetful state, while the subsequent write pulse across one of the memristors provides a learning signal to the synapse.

% Memristor Figure

Using a memristor model called the ``Metastable Switch Model'' \cite{nugent2014ahah} we have developed a simulation platform to run machine learning applications and benchmarks on top of emulated kT-RAM circuits. To date we have demonstrated state-of-the-art benchmark performance on a handful of standard machine learning classification benchmarks including the popular MNIST hand-written digits dataset (1.0\% error rate). Additionally, we have demonstrated real-time clustering, time-series signal prediction, combinatorial optimization, anomaly detection, unsupervised feature learning and reinforcement learning. Demonstrations have been previously published in ``AHaH Computing -- From Metastable Switches to Attractors to Machine Learning'' \cite{nugent2014ahah} including companion open source code (\url{https://github.com/timmolter/ahah}), a web article series titled Machine Learning Capabilities with Thermodynamic RAM and the Knowm API (\url{http://knowm.org/machine-learning-capabilities-with-thermodynamic-ram-and-the-knowm-api/}), and a play list on YouTube called Emulated Thermodynamic RAM Demonstrations (\url{https://www.youtube.com/playlist?list=PLlDmU5euIpk0qiErPtBhtrtL3_53jtnL9}). Recent advancements in deep learning from various groups have further improved the scores on ML benchmarks, and we are currently developing our own deep learning capabilities on emulated kT-RAM and tackling more challenging image-recognition benchmarks. To date we have successfully produced the memristors we need that exhibit bi-directional incremental learning as presented here. The next steps are to build the first AHaH Node prototype circuits, followed by fully-functional Thermodynamic RAM.

\section*{Acknowledgment}
The authors would like to thank the Air Force Research Labs in Rome, NY for their support under the SBIR/STTR programs AF10-BT31, AF121-049. The authors would like to thank Kristy A. Campbell from Boise State University for graciously providing us with memristor device data.

\bibliographystyle{IEEEtran}
\bibliography{2016_ML_Memristors}

\begin{thebibliography}{1}
\providecommand{\url}[1]{#1}
\csname url@samestyle\endcsname
\providecommand{\newblock}{\relax}
\providecommand{\bibinfo}[2]{#2}
\providecommand{\BIBentrySTDinterwordspacing}{\spaceskip=0pt\relax}
\providecommand{\BIBentryALTinterwordstretchfactor}{4}
\providecommand{\BIBentryALTinterwordspacing}{\spaceskip=\fontdimen2\font plus
\BIBentryALTinterwordstretchfactor\fontdimen3\font minus
  \fontdimen4\font\relax}
\providecommand{\BIBforeignlanguage}[2]{{%
\expandafter\ifx\csname l@#1\endcsname\relax
\typeout{** WARNING: IEEEtran.bst: No hyphenation pattern has been}%
\typeout{** loaded for the language `#1'. Using the pattern for}%
\typeout{** the default language instead.}%
\else
\language=\csname l@#1\endcsname
\fi
#2}}
\providecommand{\BIBdecl}{\relax}
\BIBdecl

\bibitem{nugent2014ahah}
M.~A. Nugent and M.~T. W, ``Ahah computing–-from metastable switches to
  attractors to machine learning,'' \emph{PLoS ONE}, vol.~9, p. e85175, 02
  2014.

\bibitem{mainzer2013cause}
K.~Mainzer and L.~Chua, ``The cause of complexity and symmetry breaking,''
  2013.

\bibitem{jorgensen2011evolutionary}
S.~E. J{\o}rgensen, \emph{Evolutionary Essays:: A Thermodynamic Interpretation
  of the Evolution}.\hskip 1em plus 0.5em minus 0.4em\relax Elsevier, 2011.

\bibitem{nugent2014thermodynamic}
M.~A. Nugent and M.~T. W, ``Thermodynamic-ram technology stack,'' \emph{arXiv
  preprint arXiv:1406.5633}, 2014.

\bibitem{nugent2014cortical}
------, ``Cortical processing with thermodynamic-ram,'' \emph{arXiv preprint
  arXiv:1408.3215}, 2014.

\end{thebibliography}

\end{document}